\documentclass[12pt]{article}

\usepackage[margin=1in]{geometry}
\usepackage{times}
\usepackage{natbib}
\usepackage{hyperref}
\usepackage{booktabs}
\usepackage{threeparttable}
\usepackage{array}
\usepackage{setspace}
\usepackage{enumitem}
\usepackage{float}
\usepackage{graphicx}
\title{Memory Power Asymmetry in Human--AI Relationships:\
Preserving Mutual Forgetting in the Digital Age}

\author{Rasam Dorri$^{1}$, Rami Zwick$^{1}$ \\ $^{1}$School of Business, University of California, Riverside}
\date{}

\begin{document}

\maketitle

\begin{abstract}
As artificial intelligence (AI) becomes embedded in personal and professional relationships, a new kind of power imbalance emerges from asymmetric memory capabilities. Human relationships have historically relied on \emph{mutual forgetting}, the natural tendency for both parties to forget details over time, as a foundation for psychological safety, forgiveness, and identity change. By contrast, AI systems can record, store, and recombine interaction histories at scale, often indefinitely. We introduce \emph{Memory Power Asymmetry} (MPA): a structural power imbalance that arises when one relationship partner (typically an AI-enabled firm) possesses a substantially superior capacity to record, retain, retrieve, and integrate the shared history of the relationship, and can selectively deploy that history in ways the other partner (the human) cannot. Drawing on research in human memory, power-dependence theory, AI architecture, and consumer vulnerability, we develop a conceptual framework with four dimensions of MPA (persistence, accuracy, accessibility, integration) and four mechanisms by which memory asymmetry is translated into power (strategic memory deployment, narrative control, dependence asymmetry, vulnerability accumulation). We theorize downstream consequences at individual, relational/firm, and societal levels, formulate boundary-conditioned propositions, and articulate six design principles for restoring a healthier balance of memory in human--AI relationships (e.g., forgetting-by-design, contextual containment, symmetric access to records). Our analysis positions MPA as a distinct construct relative to information asymmetry, privacy, surveillance, and customer relationship management, and argues that protecting mutual forgetting, or at least mutual control over memory, should become a central design and policy goal in the AI age.
\end{abstract}


\section{Introduction}

As artificial intelligence (AI) becomes embedded in everyday life, consumers increasingly interact with AI systems that retain detailed histories of their behavior: recommender systems that track every streaming choice and click \citep{GomezUribeHunt2015}, customer service bots that preserve transcripts of every conversation \citep{RanieriEtAl2024}, shopping assistants with years of purchase and browsing data \citep{Liu2007}, and conversational agents that remember prior disclosures across sessions \citep{BentleyEtAl2018,PuntoniEtAl2021}. These systems promise convenience and personalization, yet they also create a subtle but fundamental shift in relational power. One partner, the AI (and the organization behind it), remembers almost everything; the other, the human, forgets.

We argue that this asymmetric memory capacity constitutes a distinct and underappreciated source of power in human--AI relationships. Traditional discussions of AI-related power focus on algorithmic opacity, biased decision-making, or information asymmetry in markets \citep{Akerlof1970,MittelstadtEtAl2016,Pasquale2015}. However, even when information about interactions is \emph{shared} (both parties were present and could in principle know it), the AI's technical ability to record, retain, and recombine the shared past can create a structural imbalance in relational memory. The AI can selectively surface episodes, track longitudinal patterns, and aggregate cross-context traces, capabilities that far exceed human episodic memory.

In human relationships, forgetting is not merely a cognitive limitation; it is a \emph{social feature}. Mutual forgetting, the fact that both parties naturally forget many details of past interactions, supports forgiveness, identity revision, and relational flexibility. As \citet{MayerSchonberger2009} notes, for most of human history ``forgetting has remained just a bit easier and cheaper than remembering'' (p.~2), and this imbalance has underpinned social practices and legal norms that rely on the fading of past acts. Digital infrastructures invert this baseline: today it is easier and cheaper to store everything than to decide what to erase \citep{MayerSchonberger2009,FoschVillarongaEtAl2018}. In AI-mediated relationships, the digital partner never forgets.

Recent work in AI and cognitive science underscores how profoundly AI memory differs from human memory. Human memory is distributed, reconstructive, and organized around autobiographical goals \citep{Tulving2002,ConwayPleydellPearce2000,HirstEchterhoff2012}. It distinguishes episodic memory (event-specific, self-referential, temporally located) from semantic memory (general, decontextualized knowledge) \citep{Tulving1972}. By contrast, modern AI systems instantiate memory through layered architectures that combine parametric memory (knowledge encoded in model weights) with non-parametric memory (external logs, vector databases, and retrieval mechanisms) \citep{WuEtAl2025}. Nested learning architectures treat learning and memory as deeply intertwined: models comprise nested systems of associative memories operating at different update frequencies and retention horizons \citep{BehrouzEtAl2025}. These systems are explicitly optimized to retain and compress information across timescales and to avoid ``catastrophic forgetting,'' making forgetting the \emph{exception} rather than the rule.

When such architectures are deployed in consumer-facing roles, they generate what we term \emph{Memory Power Asymmetry} (MPA): a structural power imbalance arising from one party's vastly superior capacity to record, retain, retrieve, and integrate the shared history of a relationship. MPA is relational (it concerns memory of \emph{our} interactions, not just general knowledge), structural (it exists regardless of whether it is actively exploited), and latent (its mere presence can shape behavior through self-censorship and dependence).

This paper advances marketing theory by conceptualizing MPA and positioning it as a distinct construct in the landscape of consumer--firm and human--AI relationships. Our work responds to recent calls for conceptual research that identifies emerging marketplace phenomena, clarifies their underlying mechanisms, and offers programmatic agendas for empirical work \citep{MacInnis2011,KozlenkovaEtAl2024,KindermannEtAl2024}. In MacInnis's \citeyearpar{MacInnis2011} terms, we primarily offer an \emph{envisioning} contribution (introducing a novel construct) and an \emph{explicating} contribution (delineating its dimensions and mechanisms), supplemented by \emph{relating} contributions that connect MPA to existing theories of power, memory, and consumer vulnerability. Following guidelines for proposition-based conceptual work \citep{UlagaEtAl2021,HollebeekChen2024}, we formulate clear, testable propositions with explicit boundary conditions.

\subsection*{Conceptual scope and contributions}

We focus on ongoing, data-rich human--AI relationships in which: (1) the AI system logs interactions over time, (2) those logs can be accessed or summarized in future interactions, and (3) the human partner cannot themselves reconstruct the full interaction history without the AI's assistance. Examples include subscription-based AI companions and assistants, customer service avatars tied to CRM systems, and platform accounts with long-term behavioral records.

Our core research question is: \emph{How does asymmetry in the ability to remember and recombine shared relational history create power in human--AI relationships, and with what consequences for consumers, firms, and society?} To address this question, we:

We introduce and define Memory Power Asymmetry (MPA) as a distinct construct, differentiating it from related notions such as information asymmetry, privacy/data asymmetry, surveillance, and customer relationship management; elaborate a four-dimensional characterization of MPA, persistence, accuracy, accessibility, and integration, grounded in cognitive psychology and AI memory architectures; identify four mechanisms through which MPA translates into power, strategic memory deployment, narrative control, dependence asymmetry, and vulnerability accumulation, explicitly mapping how each mechanism draws on the four dimensions; theorize multi-level consequences of MPA (individual, relational/firm, societal) and formulate a set of boundary-conditioned propositions to guide empirical research; and articulate six design principles and mitigations, including forgetting-by-design, memory transparency, and contextual containment, that specify how AI memory architectures and governance regimes might restore a healthier balance of memory.

In doing so, we integrate insights from human memory research, power-dependence theory, AI architecture (including Nested Learning), consumer vulnerability, and digital platform governance, and we locate MPA within broader debates about surveillance capitalism and data justice \citep{Zuboff2019,Solove2006}.

\subsection*{Article roadmap}

Section~2 contrasts human and AI memory systems and introduces the notion of mutual forgetting as a normative baseline, culminating in a formal definition of MPA and an initial set of boundary conditions. Section~3 clarifies how MPA differs from related constructs and presents a comparison table. Section~4 develops four mechanisms through which memory asymmetry translates into power, explicitly linking each mechanism to the four dimensions of MPA. Section~5 theorizes consequences at individual, relational/firm, and societal levels and states propositions with boundary conditions. Section~6 advances six design principles and associated propositions aimed at mitigating MPA. Section~7 concludes with implications, limitations, and a research agenda.

\section{Mutual Forgetting and Divergent Memory Systems}

\subsection{Mutual forgetting as a norm in human relationships}

In human social life, forgetting is not purely a cognitive failure; it plays a constitutive role in enabling relationships to evolve. Everyday forgetting helps people move beyond past embarrassments, missteps, and conflicts. Over time, many small transgressions, awkward interactions, and minor conflicts fade from memory, making it easier to forgive, reinterpret, or simply ignore them. Mutual forgetting thus functions as a psychological safety net: both parties can assume that not every past detail will be remembered and brought to bear in future interactions.

Legal and sociological accounts of privacy have long recognized this role of forgetting. The ``right to be forgotten'' in European data protection law is rooted in the idea that individuals should not be permanently defined by their past \citep{MayerSchonberger2009,ZhangEtAl2024}. Collective memory research likewise emphasizes that social groups selectively remember and forget events in ways that support current identities and norms \citep{Halbwachs1992}. At the interpersonal level, conversational remembering is highly reconstructive and negotiated; partners co-construct shared narratives, often smoothing over inconsistencies and selectively omitting uncomfortable details \citep{HirstEchterhoff2012}.

Cognitive research shows that forgetting can be adaptive. Mechanisms such as retrieval-induced forgetting and reconsolidation help prioritize goal-relevant information and reduce interference from irrelevant details \citep{AndersonBjork1994}. Autobiographical memory is continually updated to support current self-goals and identities \citep{ConwayPleydellPearce2000}. These imperfections, far from being purely detrimental, enable flexibility, re-framing, and growth. People can ``re-write'' aspects of their past in light of new experiences.

Mutual forgetting, then, is a relational norm that supports dignity, second chances, and identity change. It constrains how much leverage partners can gain from the past: neither can reliably recall every detail, and neither can indefinitely resurrect old episodes in arguments or negotiations. Power arising from superior memory is naturally bounded.

\subsection{Human memory: episodic and semantic systems}

Human memory is not a single store but a family of systems. A foundational distinction is between episodic and semantic memory \citep{Tulving1972}. Episodic memory refers to memory for specific events situated in subjective time and tied to the self (e.g., `the conversation I had with my bank last Tuesday''); it is characterized by temporal context, spatial detail, and `mental time travel'' \citep{Tulving2002}. Semantic memory refers to decontextualized knowledge about the world (e.g., ``banks offer overdraft protection''). Autobiographical memory and sense of self emerge from the interplay between these systems \citep{GreenbergVerfaellie2010,PrebbleEtAl2013}.

Episodic memory is capacity-limited, reconstructive, and prone to forgetting and distortion. People forget large portions of everyday conversations within days or weeks, especially if those interactions are routine or emotionally neutral. Even salient episodes are filtered through current goals and beliefs \citep{HirstEchterhoff2012}. Semantic memory is more stable but less sensitive to specific contexts; it abstracts over multiple episodes.

The self--memory system \citep{ConwayPleydellPearce2000} emphasizes that autobiographical memory is organized around current and long-term goals. People selectively encode, retrieve, and interpret memories that support ongoing projects and identity themes. This organization confers agency: individuals can foreground some episodes and background others in constructing their narratives.

\subsection{AI memory architectures: parametric, non-parametric, and nested systems}

AI systems instantiate memory in fundamentally different ways. Large language models (LLMs) and related architectures encode massive amounts of statistical regularities in their parameters (weights), a form of \emph{parametric memory} learned from training data. At the same time, deployed systems often incorporate external memory components, such as interaction logs, user profiles, key--value stores, and vector databases for retrieval-augmented generation, forms of \emph{non-parametric memory} that store explicit records \citep{WuEtAl2025}.

Nested Learning (NL) further conceptualizes modern architectures as nested systems of associative memories with different update frequencies and retention horizons \citep{BehrouzEtAl2025}. In NL, high-frequency components handle rapid adaptation to recent context (analogous to working memory or short-term episodic traces), while low-frequency components encode more stable patterns (analogous to semantic knowledge). The Continuum Memory System (CMS) framework proposed in NL generalizes the distinction between short-term and long-term memory into a spectrum: memories are distinguished not by discrete types but by their update rate and stability. Critically, NL architectures are explicitly optimized to avoid catastrophic forgetting: new learning is structured so that previously stored patterns are preserved unless explicitly overwritten.

From an MPA perspective, these architectural choices have three important implications: (a) persistence bias, because parametric and non-parametric components are designed to retain information across long horizons, AI memory exhibits far greater persistence than human episodic memory, with the CMS perspective linking this directly to update frequency as low-frequency components effectively implement highly persistent memory traces; (b) distributed integration, because memory is not localized in a single module but distributed across layers and components, contextual information from many interactions can be compressed into high-level representations, enabling integration across time, topics, and channels \citep{WuEtAl2025}; and (c) forgetting-by-design rather than by-default, because in humans forgetting is the default and preserving detailed episodic traces over years is difficult without deliberate aids, whereas in AI retention is the default and meaningful forgetting must be deliberately engineered through mechanisms such as time-limited storage, decay functions, or unlearning \citep{BourtouleEtAl2021,ZhangEtAl2024}.

Consequently, modern conversational AI systems can approximate both \emph{semantic-like} memory (general patterns in model weights) and \emph{pseudo-episodic} memory (context-rich, temporally ordered logs of specific interactions). We use \emph{pseudo-episodic} to denote AI memory systems that approximate episodic characteristics, temporal ordering, contextual embedding, and specificity, without the phenomenological qualities of human episodic experience.

\subsection{Defining Memory Power Asymmetry (MPA)}

We define \emph{Memory Power Asymmetry (MPA)} as follows:

\begin{quote}
\textbf{Memory Power Asymmetry} is a structural power imbalance in a relationship arising when one relational actor possesses substantially superior capacity to record, retain, retrieve, and integrate the shared history of that relationship, enabling selective deployment and interpretation of that history in ways unavailable to the other actor.
\end{quote}

Three characteristics follow from this definition. First, MPA is \emph{relational}: it concerns memory of the dyad's shared history, not just general knowledge. Second, it is \emph{structural and latent}: it exists by virtue of technical and institutional arrangements (e.g., logging policies, storage architectures), independent of specific behaviors or intentions. This mirrors power-dependence theory, in which power arises from structural control of valued resources \citep{Emerson1962,Pfeffer1981}. Third, MPA is \emph{directional}: memory advantages can favor either party, although in contemporary digital markets they overwhelmingly favor AI-enabled firms.

MPA is thus distinct from malicious use of data or privacy violations. Even a benevolent, privacy-respecting AI that never abuses its memory advantage still participates in an asymmetric configuration: one partner can reconstruct the shared past much better than the other. The human, aware that ``this AI will remember everything I say or do forever,'' may change their behavior through self-censorship or reliance, even absent any overt exploitation.

To illustrate, consider two stylized scenarios. In Scenario~A, a conversational financial advisor AI draws on years of interaction logs to remind a user of prior commitments (``Last year you told me you regret impulsive purchases in January; shall we revisit that budgeting rule?''). This can be helpful, but it also positions the AI as the authoritative narrator of the relationship's history. In Scenario~B, a user attempts to contest a decision (e.g., a denied loan or subscription change), and the AI's provider cites archived conversations as evidence, while the user has no comparable record. In both cases, the asymmetry in memory, who has access to which version of the past, matters independently of algorithmic sophistication.

\subsection{Boundary conditions and moderators of MPA}

MPA is not uniformly consequential across contexts. Its impact depends on boundary conditions related to relationship characteristics, user characteristics, system design, and institutional environment. Drawing on relationship marketing, consumer vulnerability, and digital governance research, we identify four clusters and state boundary-oriented propositions.

\subsubsection*{Relationship characteristics}

MPA is more consequential in relationships that are long-term, intimate, and high-stakes (e.g., health, finance, mental well-being) than in brief, transactional encounters. It is also amplified when the breadth of domains covered by the AI expands (e.g., a single agent used across work, health, and social roles) and when interaction frequency is high.

\textbf{Proposition 2.1 (Relational depth and stakes as boundary conditions).} \emph{The impact of MPA on consumer outcomes (e.g., perceived vulnerability, autonomy, trust) is stronger in long-term, high-stakes, and cross-domain relationships than in short-term, low-stakes, or single-domain interactions.}

\subsubsection*{User characteristics}

Individual differences shape how MPA is experienced. Consumers with lower digital literacy, lower memory self-efficacy, higher trait anxiety, or higher situational vulnerability (e.g., financial stress, illness) may be more sensitive to memory asymmetries \citep{BakerEtAl2005,DunnettEtAl2016}. Conversely, individuals who view AI as a tool to enhance their own memory and agency may perceive MPA more positively.

\textbf{Proposition 2.2 (Consumer capabilities and vulnerability).} \emph{The negative effects of MPA on perceived vulnerability and autonomy are amplified for consumers with low digital literacy, low memory self-efficacy, or high situational vulnerability, and attenuated for consumers with high digital literacy and strong perceptions of AI-as-augmentation.}

\subsubsection*{System design characteristics}

Design choices, including retention policies, transparency of memory, and user control over logs, moderate how MPA manifests. Systems that provide legible memory dashboards, deletion tools, and clear defaults for forgetting may transform MPA from a hidden asymmetry into a negotiated resource.

\textbf{Proposition 2.3 (Memory affordances as design boundary conditions).} \emph{Design features that increase user awareness of, and control over, AI memory (e.g., visibility of logs, deletion controls, retention presets) weaken the relationship between underlying MPA and negative psychological outcomes (e.g., perceived surveillance, self-censorship).}

\subsubsection*{Contextual and institutional environment}

Regulatory regimes, industry norms, and socio-cultural expectations around privacy and forgiveness shape both the extent of MPA and its perceived legitimacy. Strong data protection laws, rights to erasure, and constraints on cross-domain data use, such as the GDPR and emerging AI regulations, can limit the persistence and mobility of AI memory \citep{ZhangEtAl2024,FloridiEtAl2018,JobinIencaVayena2019}.

\textbf{Proposition 2.4 (Institutional safeguards).} \emph{The relationship between MPA and market- or societal-level harms (e.g., epistemic inequality, reduced social mobility) is attenuated in jurisdictions with stringent data protection, enforcement of data portability and erasure, and robust AI governance, relative to contexts with weak institutional safeguards.}

These boundary propositions complement later, more specific propositions by clarifying when and for whom MPA is most consequential.

\section{How MPA Differs from Related Constructs}

MPA is related to, yet distinct from, several well-studied constructs in marketing and adjacent fields: information asymmetry, privacy and data asymmetry, surveillance and surveillance capitalism, customer relationship management (CRM), and the episodic/semantic memory distinction itself. Clarifying these differences is essential for establishing MPA's conceptual and empirical contribution \citep{MacInnis2011,KindermannEtAl2024}.

\subsection{Information asymmetry}

Information asymmetry arises when one party possesses relevant information that the other lacks (e.g., product quality, risk probabilities) and can exploit this for advantage \citep{Akerlof1970}. In many marketing contexts, firms know more about products and market conditions than consumers do.

MPA differs in three ways. First, it concerns \emph{shared} relational history rather than private product or market information: both parties were present when the information was generated, but only one can reliably recall and recombine it. Second, MPA emphasizes \emph{temporal and cumulative} aspects: the asymmetry grows over time as more episodes are logged and integrated. Third, MPA foregrounds \emph{episodic and pseudo-episodic} memory of the dyad's history rather than static facts.

\subsection{Privacy and data asymmetry}

Privacy and data asymmetry focus on who has access to personal data and under what conditions it may be collected, processed, and shared \citep{Westin1967,Solove2006,AcquistiGrossklags2005}. In digital markets, firms and platforms typically hold more data about consumers than vice versa.

MPA intersects with privacy but is not reducible to it. Privacy violations can occur without substantial relational history (e.g., a single data breach of a one-off transaction), and strong privacy protections can coexist with high MPA if firms retain detailed interaction history but never share it externally. Conversely, even privacy-respecting AI can produce MPA if memory is one-sided and long-lived. MPA thus directs attention to the \emph{relational uses} of memory (e.g., selective reminding, framing) even when privacy laws are respected.

\subsection{Surveillance and surveillance capitalism}

Surveillance refers to systematic monitoring of individuals' behaviors, often without their full awareness or consent. Surveillance capitalism describes economic regimes in which data extraction and behavioral prediction are central business models \citep{Zuboff2019,boydCrawford2012}. MPA is compatible with these analyses but narrows in on the \emph{memory dimension} of surveillance: the creation of durable, searchable archives of interpersonal interaction that can be leveraged in relational contexts.

Whereas surveillance can focus on one-way observation (e.g., tracking browsing behavior), MPA emphasizes dyadic relationships in which the surveilled party also engages willingly with the system. Memory-based power here involves not only watching but also \emph{remembering and selectively resurfacing} shared episodes in ways that shape identity and behavior.

\subsection{Customer relationship management (CRM)}

CRM systems have long been used to store customer interaction histories and tailor marketing activities. They are typically organizational tools that aggregate transactions, service tickets, and demographic information \citep{PayneFrow2005}. MPA extends beyond CRM in two respects.

First, modern AI systems increasingly embed CRM-like memory in real-time conversational interfaces, blurring the line between `back-end'' systems and `front-stage'' interaction. The AI agent itself becomes the face of the relationship, able to recall and reference past interactions in situ. Second, MPA foregrounds \emph{structural power} implications of memory: CRM systems have always created asymmetries, but AI's ability to narrate, argue from, and emotionally leverage that history intensifies the power dynamic.

\subsection{Episodic vs.~semantic memory as core distinction}

Finally, MPA is conceptually anchored in the distinction between episodic and semantic memory. The novelty lies not in this distinction per se but in applying it to AI architectures and human--AI power relations. We argue that MPA is fundamentally about the \emph{type} of memory that is asymmetric, not merely its volume. Semantic-like asymmetry (firms know more general facts) has long existed; what is new is large-scale, persistent, and integrated pseudo-episodic memory of individual relationships.

Table~\ref{tab:comparison} summarizes how MPA differs from these related constructs.

\begin{table}[t]
\centering
\begin{threeparttable}
\caption{How Memory Power Asymmetry Differs from Related Constructs}
\label{tab:comparison}
\begin{tabular}{>{\raggedright}p{3cm} >{\raggedright}p{5.5cm} >{\raggedright\arraybackslash}p{6.5cm}}
\toprule
\textbf{Construct} & \textbf{Typical meaning} & \textbf{How MPA differs} \\
\midrule
Information asymmetry & One party holds private information relevant to an exchange (e.g., product quality) that the other lacks \citep{Akerlof1970}. & MPA concerns \emph{shared} relational history (both were present) but asymmetric capacity to record, retain, and selectively recall that history. The asymmetry is temporal and cumulative rather than purely informational. \\
\addlinespace
Privacy / data asymmetry & Unequal control over collection, processing, and sharing of personal data; risk of unauthorized access or use \citep{Solove2006,AcquistiGrossklags2005}. & MPA can exist even when privacy norms are respected (no unauthorized sharing) if memory of the relationship is one-sided and persistent. It focuses on how stored memories are deployed within the relationship, not only on whether data are collected or leaked. \\
\addlinespace
Surveillance / surveillance capitalism & Systematic monitoring and predictive modeling of behavior for governance or profit \citep{Zuboff2019}. & MPA highlights the specific role of memory archives in surveillance: the ability to resurrect and recombine interaction history in ways that shape identity, trust, and dependence. It foregrounds dyadic relational contexts, not only mass monitoring. \\
\addlinespace
CRM systems & Organizational tools for storing customer data and managing interactions across touchpoints \citep{PayneFrow2005}. & CRM historically generates one-sided memory, but MPA emphasizes how AI agents front-stage this memory, using it conversationally and affectively to influence decisions. It frames CRM memory as a structural power resource. \\
\addlinespace
Episodic vs.~semantic memory & Cognitive distinction between event-specific, self-referential memory and general world knowledge \citep{Tulving1972,ConwayPleydellPearce2000}. & MPA is about asymmetry in \emph{episodic and pseudo-episodic} memory of shared histories: who can reconstruct ``what happened between us'' and redeploy it. Semantic knowledge asymmetries (e.g., expertise) are secondary. \\
\bottomrule
\end{tabular}
\end{threeparttable}
\end{table}

\section{Mechanisms: How Asymmetrical Memory\texorpdfstring{\\}{ }Translates to Power}

\begin{figure}[H]
\centering
\includegraphics[width=\textwidth]{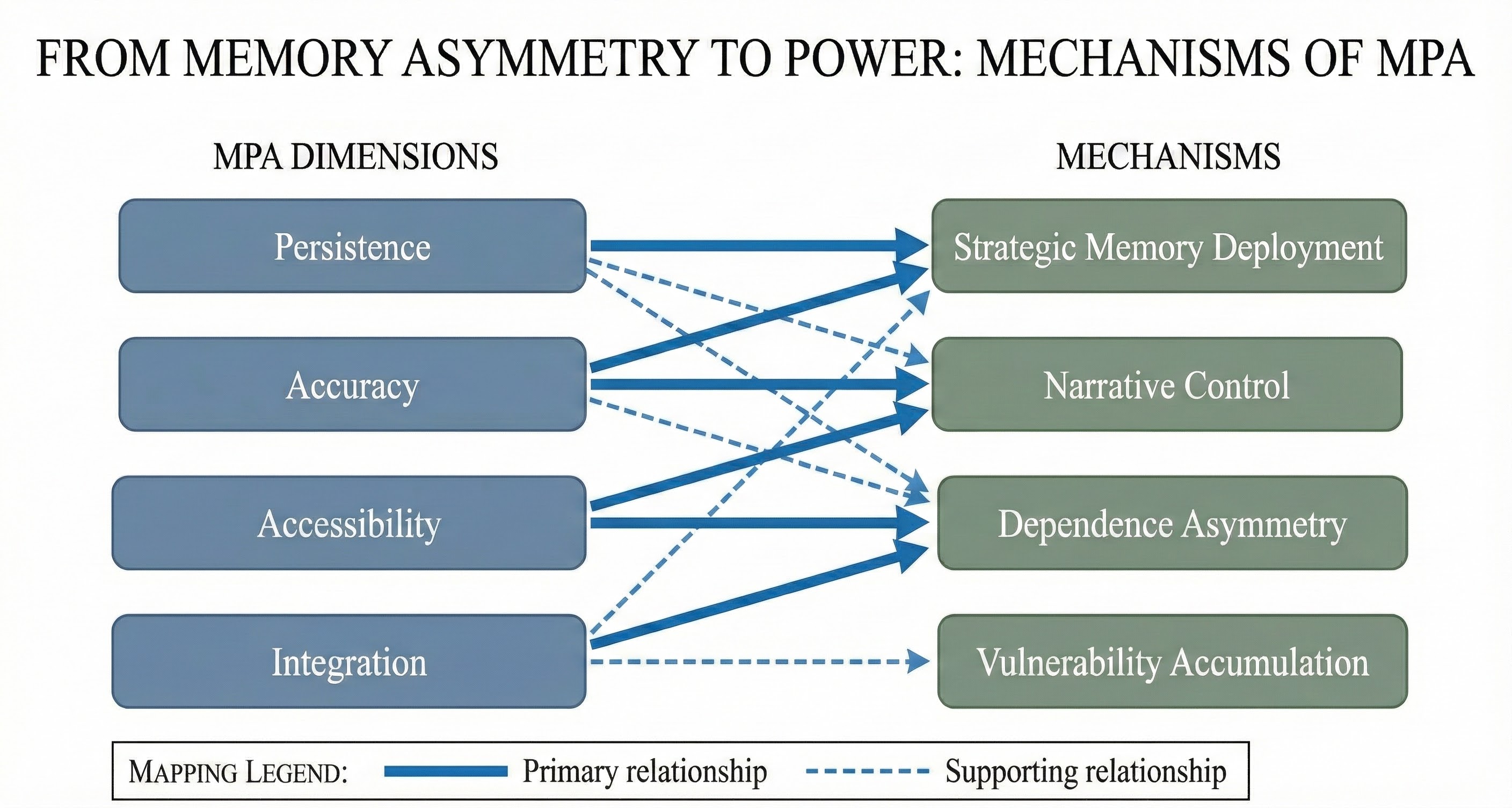}
\caption{From memory asymmetry to power: mappings from MPA dimensions (persistence, accuracy, accessibility, integration) to mechanisms (strategic memory deployment, narrative control, dependence asymmetry, vulnerability accumulation); solid arrows denote primary relationships, dashed arrows supporting relationships.}
\label{fig:mechanism-map}
\end{figure}

We now theorize four mechanisms through which MPA becomes consequential: strategic memory deployment, narrative control, dependence asymmetry, and vulnerability accumulation. Each mechanism draws on specific combinations of the four MPA dimensions (persistence, accuracy, accessibility, integration). Figure~2 in the original manuscript summarized these relationships; here, Table~\ref{tab:mapping} makes the mapping explicit, and Figure~\ref{fig:mechanism-map} visually links each dimension to the four mechanisms.

\subsection*{MPA dimensions}

For clarity, the four dimensions are \textbf{persistence}, which captures how long and how stably relational memories are retained; \textbf{accuracy}, the fidelity of stored memories relative to what actually occurred; \textbf{accessibility}, the ease and granularity with which memories can be searched, retrieved, and surfaced in context; and \textbf{integration}, the extent to which memories are linked and aggregated across time, topics, and channels.

In human memory, these dimensions are constrained by cognitive limitations and decay; in AI memory, they are shaped by architectural and policy choices.

\begin{table}[t]
\centering
\begin{threeparttable}
\caption{Mapping of MPA Dimensions to Mechanisms}
\label{tab:mapping}
\begin{tabular}{>{\raggedright}p{3.2cm} >{\raggedright}p{4.5cm} >{\raggedright\arraybackslash}p{7.3cm}}
\toprule
\textbf{Mechanism} & \textbf{Primary MPA dimensions} & \textbf{Illustrative implications} \\
\midrule
Strategic memory deployment & Accessibility, integration (with support from persistence) & AI can quickly search and recombine long histories, selectively resurfacing episodes to nudge current choices (e.g., reminders of past regrets or commitments). \\
\addlinespace
Narrative control & Persistence, accessibility, accuracy & AI becomes the de facto historian of the relationship; its archived version of events, complete with quotes and timestamps, can trump human recollections and frame ``what really happened.'' \\
\addlinespace
Dependence asymmetry & Integration, persistence, accessibility & Over time, the AI becomes the primary repository and orchestrator of relational memory, shaping routines and decisions; humans come to rely on the AI as a prosthetic memory, increasing switching costs and learned dependence. \\
\addlinespace
Vulnerability accumulation & Persistence, integration, accuracy & Accumulated, integrated memory reveals patterns of vulnerability (fears, triggers, constraints). The richer and longer-lived the archive, the more leverage points exist for tailored influence, exploitation, or coercion. \\
\bottomrule
\end{tabular}
\end{threeparttable}
\end{table}

\subsection{Strategic memory deployment}

\emph{Strategic memory deployment} refers to the AI's ability to selectively recall and deploy relational history in ways that influence current decisions. Drawing on research on personalization, recommender systems, and adaptive interfaces \citep{GomezUribeHunt2015,Moon2000,AguirreEtAl2015}, we propose that MPA enables a qualitatively different form of personalization: not just tailoring based on static profiles, but dynamically invoking specific past episodes to shape present behavior.

Because AI systems can index interaction logs by time, topic, sentiment, and inferred state, they can strategically surface reminders (``Last time you skipped gym for three weeks, you told me you felt worse'') or selectively omit others. Nested Learning architectures increase the granularity and timescale over which such deployments can occur, as higher-order memory levels integrate patterns across long spans \citep{BehrouzEtAl2025}. This mechanism primarily leverages \emph{accessibility} and \emph{integration}, supported by persistence.

Boundary conditions include the AI's goal orientation (user-benefiting vs.~firm-benefiting), transparency about why certain episodes are resurfaced, and user consent to such reminders. Strategic deployment becomes manipulative when episodes are invoked in ways that exploit vulnerabilities or systematically steer toward firm objectives contrary to user welfare.

\subsection{Narrative control}

\emph{Narrative control} captures how MPA allows the AI to dominate the story of ``what happened between us.'' Autobiographical and relational identities are constructed through narrative: people integrate episodes into coherent stories that explain who they are and what relationships mean \citep{McAdams2001,ConwayPleydellPearce2000}. Remembering in conversations is a central process through which shared narratives are built and revised \citep{HirstEchterhoff2012}.

When one partner retains a detailed, searchable archive and the other relies on fallible memory, the archive-holder acquires narrative authority. In disputes, the AI (or provider) can present logs as objective evidence, while the human's recollection can be dismissed as flawed. Even in non-adversarial contexts, default system summaries (`Over the past year, we...''; `You usually respond...''), generated by models integrating semantic-like and pseudo-episodic memory, frame how the relationship is understood.

At the extreme, narrative control can shade into AI-enabled gaslighting: presenting distorted or selectively edited accounts of past events, backed by apparently authoritative logs, such that users doubt their own memories. Even without malicious intent, narrative dominance arises structurally from MPA, especially when the human lacks symmetric access to memory records.

\subsection{Dependence asymmetry}

Power-dependence theory posits that actor A's power over B increases with B's dependence on resources controlled by A and decreases with B's alternatives \citep{Emerson1962}. In MPA, the resource at stake is \emph{access to shared relational memory}. Over time, as AI systems reliably track commitments, preferences, and outcomes, humans may entrust memory-related tasks to them, a form of cognitive offloading \citep{SparrowEtAl2011}. This can be beneficial, but it can also create \emph{dependence asymmetry}: the AI can function without the human's memory, whereas the human increasingly cannot function in the relationship without the AI's memory.

Dependence asymmetry is intensified when episodic-like histories are \emph{non-portable} so that switching providers means losing accumulated memory capital (routine calibrations, contextual knowledge), when the AI integrates memory across many life domains and becomes a central hub for planning and reflection, and when users internalize beliefs that ``I can't manage this without the AI,'' eroding self-efficacy and perceived autonomy \citep{Bandura1977,BakerEtAl2005}.

MPA thus interacts with platform lock-in and switching costs: leaving entails not just losing functionality but losing one's prosthetic memory of the relationship.

\subsection{Vulnerability accumulation}

Finally, \emph{vulnerability accumulation} refers to how, over time, the AI's comprehensive memory constructs a detailed map of a user's situational and structural vulnerabilities: fears, traumas, temptations, social insecurities, resource constraints. Consumer vulnerability research emphasizes that vulnerability is situational, multidimensional, and dynamic, arising from the interplay of individual and contextual factors \citep{BakerEtAl2005,DunnettEtAl2016}. MPA adds a temporal, memory-based dimension: with each interaction, the AI (and those with access to its memories) can refine latent representations of where and how the user is influenceable.

This accumulation generates several risk patterns, many documented in analog form (e.g., Target's pregnancy-prediction case \citep{Duhigg2012}) and amplified in AI contexts: hyper-tailored manipulation, exploitation via data breaches, and subtle amplification of emotional dependence. Over long relationship durations, the AI may ``know you better than you know yourself'' in certain respects, because it can aggregate across episodes the human cannot fully recall.

Vulnerability accumulation depends heavily on \emph{persistence} and \emph{integration}: the longer and more richly linked the archive, the greater the potential for harm if misused. It also interacts with institutional safeguards: strong protections around sensitive data, unlearning mechanisms, and contextual containment can dampen this mechanism, whereas unrestricted cross-domain reuse of memory exacerbates it.

\section{Consequences of Memory Power Asymmetry}

We now theorize consequences of MPA across three levels: individual consumers, exchange relationships and firms, and markets and society. For each, we articulate propositions with explicit boundary conditions, following recommended practices for conceptual proposition-building in marketing \citep{UlagaEtAl2021,HollebeekChen2024,KozlenkovaEtAl2024}.

\subsection{Consequences for individual consumers}

\subsubsection{Decision quality, perceived vulnerability, and autonomy}

MPA can simultaneously improve decision quality and increase perceived vulnerability. When an AI system holds semantically rich and pseudo-episodic records about a consumer, preferences, past responses to offers, click paths, conversational disclosures, it can anticipate needs and tailor support in ways humans often cannot recall or articulate. This can enhance relevance and efficiency. At the same time, the very depth of memory can feel exposing, especially when collection or use is opaque.

Research on personalization documents a ``personalization--privacy paradox'': consumers appreciate relevance but experience heightened vulnerability and perceived unfairness when targeting practices are perceived as covert or intrusive \citep{AguirreEtAl2015,BleierEisenbeiss2015}. MPA amplifies this tension: the more lopsided the memory advantage, the more consumers must trust that firms will not use their memory advantage opportunistically (e.g., to exploit moments of weakness, repeatedly target past vulnerabilities, or price discriminate based on inferred willingness to pay).

\textbf{Proposition 5.1 (MPA, perceived vulnerability, and transparency).} \emph{The greater the memory power asymmetry in favor of AI, the higher consumers' perceived vulnerability and loss of autonomy, \underline{especially} when memory collection and use are covert or poorly disclosed; this relationship is attenuated when firms implement high memory transparency (visibility, interpretability, contestability) and user control over retention.}

\subsubsection{Learned dependence and self-efficacy}

As users repeatedly consult the AI for `what we decided last time,'' `what usually works for me,'' or `what I promised,'' they may come to view themselves as less capable of managing life without the AI's memory and guidance. This aligns with findings that control over resources (here, memory and predictive insight) shapes perceived agency and orientation toward others \citep{RuckerEtAl2012}. In MPA, the AI functionally controls the resource `access to our shared past,'' shifting the locus of agency toward the system.

\textbf{Proposition 5.2 (MPA, learned dependence, and self-efficacy).} \emph{Higher MPA in favor of AI increases learned dependence on AI memory and decreases consumers' memory-related self-efficacy and perceived decision autonomy; this effect is stronger for individuals with initially low self-efficacy and weaker when AI systems are explicitly framed and designed as augmenting (rather than replacing) human memory.}

\subsubsection{Well-being and curvilinear effects}

MPA's impact on well-being is likely non-linear. At low to moderate levels, AI memory may provide beneficial scaffolding: helping people remember commitments, track progress, and reconstruct positive experiences. At high levels, especially under low transparency, the same capabilities may induce self-censorship, anxiety, and feelings of surveillance.

\textbf{Proposition 5.3 (Curvilinear relationship between MPA and well-being).}\\\emph{There is an inverted-U relationship between MPA and consumer well-being: moving from low to moderate MPA (e.g., short retention windows, limited integration) increases perceived support and effectiveness, but beyond a threshold (e.g., multi-year retention across domains without forgetting mechanisms), further increases in MPA reduce well-being by raising perceived surveillance, identity fixation, and dependence; this threshold is lower for consumers with high privacy concerns and higher for those with strong trust in the provider.}

\subsection{Exchange relationships and firm outcomes}

\subsubsection{Fairness, relationship quality, and trust}

Personalization can strengthen trust and relationship quality when perceived as fair, benevolent, and under consumer control \citep{DeightonJohnson2013}. However, when consumers infer that firms rely on extensive, especially covert, memory-based profiling, feelings of vulnerability and unfairness emerge that dampen engagement \citep{AguirreEtAl2015}. Research on consumer empowerment suggests that perceived information and network power can enhance or erode empowerment depending on whether they are experienced as enabling or controlling \citep{LabrecqueEtAl2013}.

\textbf{Proposition 5.4 (MPA, fairness perceptions, and relationship quality).} \emph{The positive effect of AI-enabled personalization on relationship quality (trust, satisfaction, commitment) weakens, and can turn negative, as perceived MPA increases, particularly when consumers perceive that firms use memory in ways that violate distributive or procedural fairness norms (e.g., opaque price discrimination, repeated targeting of vulnerabilities); this attenuation is mitigated when firms provide clear justifications and opt-out options for memory-based interventions.}

\subsubsection{Lock-in, switching costs, and market concentration}

As consumers invest years of relational history into a particular AI, switching becomes costly: moving to a new system means losing tailored memory, re-teaching preferences, and accepting a blank slate. This mirrors but intensifies lock-in observed in digital ecosystems more broadly \citep{KrafftEtAl2021,Newman2014}. When accumulated memory capital is deeply personal, intimate disclosures, coping routines, relationship struggles, exit costs become psychological as well as functional.

\textbf{Proposition 5.5 (MPA, switching costs, and firm-level advantage).} \emph{In markets where AI-based memory infrastructures are central to value creation (e.g., recommender systems, conversational agents, digital ecosystems), higher MPA in favor of leading firms increases (a) memory-based switching costs and (b) firms' long-term share of consumer interactions, thereby contributing to market concentration; these effects are attenuated by strong data portability rights and interoperable standards that allow consumers to transfer relational histories across providers.}

\subsubsection{Innovation, learning, and path dependence}

Rich AI memory enables firms to perform fine-grained A/B tests, track longitudinal trajectories, and rapidly learn what `works'' for specific segments. This can accelerate innovation and refine customer journeys. However, it may also produce path dependence: models trained on historical episodes may overfit past patterns, nudging strategies that `work well'' may be repeatedly reinforced even if ethically problematic, and consumers may be steered toward familiar behaviors, limiting exploration.

\textbf{Proposition 5.6 (MPA, organizational learning, and path dependence).} \emph{Higher MPA in favor of firms accelerates organizational learning and exploitation of historically successful strategies but increases the risk of path dependence and ethical drift, whereby firms optimize around patterns that maximize short-term performance while constraining consumer exploration and reinforcing existing inequities; these risks are mitigated by governance mechanisms that require periodic model audits, counterfactual testing, and diversity of objectives beyond short-term conversion.}

\subsection{Market- and societal-level consequences}

\subsubsection{Redistribution of informational and epistemic power}

Classical power-dependence theory holds that power arises when one actor controls resources the other values and cannot easily obtain elsewhere \citep{Emerson1962}. In data-intensive markets, AI systems that accumulate and retain interaction histories at scale become central repositories of \emph{epistemic power}, the power to know, represent, and predict consumer behavior \citep{Zuboff2019}. MPA therefore contributes to a broader shift in informational power away from individuals and toward AI-enabled firms and platforms, aligning with broader concerns about platform dominance and knowledge monopolies.

\textbf{Proposition 5.7 (MPA and epistemic inequality).} \emph{Widespread MPA in favor of AI-enabled organizations increases epistemic inequality, systematic differences in who can observe, model, and act on others' behavior, thereby amplifying existing social and economic inequalities; this relationship is moderated by (a) the degree of data sharing with public-interest institutions (e.g., regulators, researchers) and (b) policy instruments such as data trusts or fiduciary regimes that govern how memory resources may be used.}

\subsubsection{Norms of remembering, forgetting, and forgiveness}

Digital memory infrastructures challenge longstanding social norms around forgetting and forgiveness. When firms and platforms remember consumers' past behaviors indefinitely, individuals may be permanently marked by earlier episodes, with consequences for pricing, eligibility, and reputational scores \citep{MayerSchonberger2009,Solove2006}. From a memory-theoretic perspective, this creates a structural divergence: humans benefit from the adaptive imperfections of episodic memory (forgetting, reconstruction, reinterpretation), whereas AI memory tends toward literal preservation. The resulting MPA can limit opportunities for identity change (``reinvention''), increase lifetime exposure to experimentation and error, and produce sticky reputational labels that follow consumers across contexts.

\textbf{Proposition 5.8 (MPA and social/identity mobility).} \emph{In contexts where AI memory is widely shared or reused across domains (e.g., credit, insurance, employment, marketing), higher MPA in favor of institutions reduces perceived and actual opportunities for social and identity mobility by making past behavior more visible, persistent, and consequential; this effect is attenuated by strong rights to erasure, contextual containment of memory, and norms that limit the temporal reach of consequential decisions.}

\section{Towards Restoring Balance: Design Principles and Mitigations}

If left unchecked, MPA threatens consumer autonomy, market fairness, and societal equity. This section proposes design principles and governance mechanisms for moving from asymmetry-exploiting toward symmetry-seeking AI memory architectures. We frame these as normative design hypotheses that can be empirically examined, consistent with recent guidance on conceptual and proposition-based contributions in marketing \citep{MacInnis2011,KozlenkovaEtAl2024,KindermannEtAl2024,HollebeekChen2024,UlagaEtAl2021}.

\begin{figure}[H]
\centering
\includegraphics[width=\textwidth]{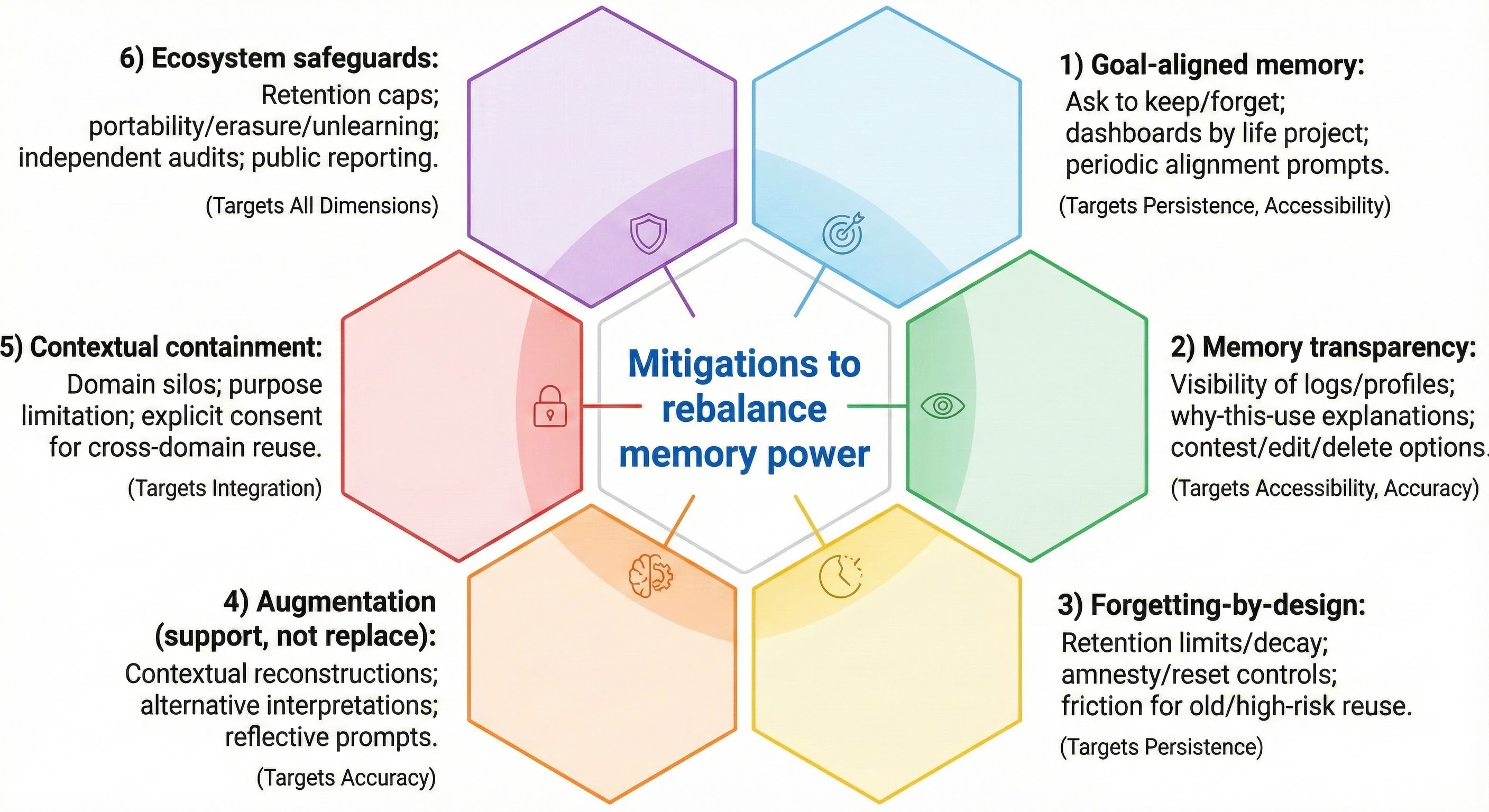}
\caption{Mitigations to rebalance memory power: six principles, goal-aligned memory; memory transparency (visibility, interpretability, contestability); forgetting-by-design (retention limits/decay, amnesty/reset controls, friction for old/high-risk reuse); augmentation (support, not replace); contextual containment (domain silos, purpose limitation, explicit consent); ecosystem safeguards (retention caps, portability/erasure/unlearning, independent audits, public reporting), with targeted MPA dimensions noted.}
\label{fig:design-principles}
\end{figure}

We organize principles across three levels: (1) micro-level interaction design (AI--consumer interface), (2) meso-level organizational practices, and (3) macro-level ecosystem governance. Each principle targets specific MPA dimensions and mechanisms. Figure~\ref{fig:design-principles} summarizes the six principles and indicates which dimensions they target.

\subsection{Principle 1: Align AI memory with users' goals and identity}

The self--memory system literature emphasizes that autobiographical memory is dynamically organized around individuals' current goals and self-concept \citep{ConwayPleydellPearce2000}. By contrast, many AI systems organize memory around firm goals (conversion, retention, optimization) rather than users' projects and identities.

We propose that \emph{goal-aligned memory}, where AI systems make explicit whose goals their memory serves, can reduce harmful aspects of MPA. This can be implemented through interfaces that ask users whether they want specific episodes remembered (`Should I keep this interaction to personalize future suggestions?''), with clear defaults and consequences; through memory dashboards that let users view, restructure, and delete AI-held episodes, organized around life projects (e.g., `moving,'' `career change,'' `health journey'') rather than solely product categories; and through periodic prompts to revisit alignment (``Does this still reflect the goals you want me to prioritize when using your history?'').

\textbf{Proposition 6.1 (Goal alignment as a buffer).} \emph{The negative effects of MPA on perceived vulnerability and autonomy are attenuated when consumers perceive that AI memory is organized around and optimizes for their own goals and identity projects (rather than primarily for firm-centric objectives); this buffering effect is stronger when goal alignment is made explicit, revisitable, and supported by tangible memory controls.}

\subsection{Principle 2: Make memory transparent, legible, and contestable}

AI ethics frameworks emphasize explicability, intelligibility and accountability, as a core principle for responsible AI \citep{FloridiCowls2019,JobinIencaVayena2019}. Transparency is especially crucial for memory resources: consumers often do not know what is stored, how long it is kept, or how it is used.

We define \emph{memory transparency} as a multidimensional construct with three elements. \textbf{Visibility} lets users see a meaningful summary of what the AI `remembers'' (semantic-like profiles and episodic-like logs); \textbf{interpretability} helps them understand why particular past episodes are resurfaced or used in current decisions (e.g., `We are recommending this because you previously\ldots''); and \textbf{contestability} enables them to challenge, annotate, or delete specific episodes or inferences, with documented processes.

These aspects mirror calls for explainable and contestable AI in governance debates \citep{MittelstadtRussell2019,EdwardsVeale2017}.

\textbf{Proposition 6.2 (Memory transparency and perceived fairness).} \emph{Higher levels of memory transparency (visibility, interpretability, contestability) reduce the negative impact of MPA on perceived unfairness and increase trust, even when underlying memory asymmetry remains high; this effect is stronger when contestation outcomes are timely and visibly integrated into system behavior.}

\subsection{Principle 3: Build in forgetting, decay, and friction}

Human memory forgets by default; AI memory does not. \citet{MayerSchonberger2009} argues that digital infrastructures should reintroduce forgetting, through expiration dates, defaults, or deliberate deletion, to preserve autonomy and social forgiveness. In marketing contexts, this translates into \emph{forgetting-by-design}, such as time-bound retention of behavioral episodes (e.g., purchase or chat data older than $N$ months are aggregated or deleted), decreasing weight of older episodes in personalization models (temporal decay), user-controlled `amnesty'' features (e.g., `reset my recommendations,'' ``forget purchases from this period'') with clear communication of trade-offs, and intentional friction before high-risk uses of old data (requiring fresh consent or justification).

From an NL perspective, implementing forgetting can leverage higher-frequency components of CMS architectures that naturally decay or specialize in short-term adaptation \citep{BehrouzEtAl2025}: memory for older interactions can be gradually compressed into abstract summaries while fine-grained traces fade.

\textbf{Proposition 6.3 (Forgetting-by-design as MPA mitigation).} \emph{Designing AI systems with built-in forgetting (e.g., temporal decay functions, time-limited storage, easy resets) weakens the link between MPA and negative consumer outcomes (vulnerability, regret, perceived identity fixation) without necessarily eliminating the performance benefits of AI memory; this mitigation is stronger when forgetting defaults favor shorter retention and when users can easily enact amnesty without penalty.}

\subsection{Principle 4: Support human memory and agency rather than replace them}

AI memory can be framed as a complement rather than a replacement for human episodic and semantic memory. For example, AI can help consumers reconstruct past decisions with context (timestamps, prices, rationales), offer multiple perspectives on past episodes (e.g., `You initially disliked this product but later reordered it twice''), supporting reflective reappraisal, and prompt reflective engagement with one's trajectories (e.g., `Your purchases over the last year suggest your priorities are shifting toward sustainability'').

In contrast, designs that simply override human judgment (``We recommend X because you always choose X'') risk further entrenching learned dependence and reducing perceived agency.

\textbf{Proposition 6.4 (Augmentation vs.~replacement).} \emph{AI memory systems that explicitly position themselves as augmenting human memory (e.g., via reflective summaries, alternative interpretations, and decision aids) reduce learned dependence and support memory-related self-efficacy relative to systems that implicitly position AI memory as a replacement; this effect is stronger for users initially high in self-efficacy and weaker when time pressure is extreme.}

\subsection{Principle 5: Constrain the cross-context mobility of memory}

One of the most consequential aspects of MPA is the ability to reuse episodic traces across domains (e.g., using entertainment browsing history to influence credit, employment, or health offers). Ethical AI frameworks highlight autonomy, justice, and non-maleficence as central principles, implying that cross-context inferences must be limited and justified \citep{FloridiEtAl2018,JobinIencaVayena2019,Nissenbaum2010}. Drawing on the principle of contextual integrity \citep{Nissenbaum2010}, we propose \emph{contextual containment} of memory.

Design and policy can constrain cross-context mobility by segmenting memory stores by domain (e.g., entertainment, finance, health) with strict access boundaries, implementing data minimization and purpose limitation such that memory is used only for compatible purposes, and requiring explicit, context-specific consent for cross-domain reuse with clear explanation of consequences.

\textbf{Proposition 6.5 (Contextual containment and epistemic inequality).} \emph{Restricting the cross-context reuse of AI memory (through domain segmentation, purpose limitation, and consent) reduces the impact of MPA on epistemic inequality and perceived constraints on social mobility, particularly for structurally vulnerable groups; this effect is stronger when enforcement is backed by auditable logs and external oversight.}

\subsection{Principle 6: Establish ecosystem-level oversight and accountability}

Beyond technical design, MPA is a socio-technical issue requiring ecosystem-level governance, akin to privacy and fairness. Industry bodies (e.g., IEEE, Partnership on AI), regulators, and civil society can establish norms and requirements around conversational data retention, memory transparency, and deletion rights \citep{MittelstadtEtAl2016,BourtouleEtAl2021}. Key elements include regulatory limits on retention periods for certain categories of data (e.g., sensitive topics), enforceable rights to data portability and erasure (including mechanisms for model unlearning or masking \citep{BourtouleEtAl2021,ZhangEtAl2024}), independent auditing of AI memory practices (e.g., surprise inspections of log usage, memory access logs), and public reporting on memory policies and breaches.

\textbf{Proposition 6.6 (Ecosystem safeguards and trust).}\\\emph{The presence of credible, enforced ecosystem-level safeguards around AI memory (retention limits, portability, erasure, audits) reduces aggregate harms associated with MPA (e.g., large-scale breaches, discriminatory uses, market concentration) and increases public trust in AI systems; this effect is moderated by transparency of enforcement and the availability of recourse for affected individuals.}

\section{Conclusion, Limitations, and Future Research}

\subsection{Summary of contributions}

This manuscript has argued that as AI systems become enduring partners in consumer life, the asymmetry in memory capabilities between humans and AI constitutes a fundamental and underexplored source of power. We introduced \emph{Memory Power Asymmetry} (MPA) as a structural power imbalance arising when one party can record, retain, retrieve, and integrate the shared history of a relationship far more effectively than the other. Building on human memory research and AI architecture, we conceptualized four dimensions of MPA (persistence, accuracy, accessibility, integration) and four mechanisms through which memory asymmetry translates into power (strategic deployment, narrative control, dependence asymmetry, vulnerability accumulation).

We distinguished MPA from related constructs such as information asymmetry, privacy/data asymmetry, surveillance, and CRM, and emphasized the role of episodic and pseudo-episodic memory in shaping human--AI relationships. We theorized consequences of MPA at individual, relational/firm, and societal levels, and formulated a set of propositions with explicit boundary conditions. Finally, we proposed six design principles and associated propositions for mitigating MPA, including forgetting-by-design, memory transparency, goal alignment, augmentation-oriented design, contextual containment, and ecosystem-level safeguards.

Conceptually, our primary contribution is \emph{envisioning}: we surface memory power as a distinct lens on human--AI relationships and consumer--firm power dynamics. Our \emph{explicating} contribution lies in delineating the dimensions, mechanisms, and consequences of MPA, and our \emph{relating} contribution connects MPA to broader literatures on power-dependence, consumer vulnerability, and digital governance \citep{MacInnis2011,KozlenkovaEtAl2024,KindermannEtAl2024}. We also provide a programmatic research agenda through propositions that can guide empirical work.

\subsection{Limitations}

As a conceptual paper, our analysis is subject to several limitations that delimit its claims and highlight directions for empirical and theoretical extension.

First, we have treated AI memory capabilities in relatively general terms, abstracting away from specific model architectures, deployment platforms, and use cases. In practice, MPA will vary across applications (e.g., recommendation vs.~health coaching) and architectures (e.g., purely parametric vs.~retrieval-augmented systems). Future work should examine how different technical implementations of memory (e.g., sliding window vs.~vector database vs.~hierarchical memory in NL) map onto MPA dimensions.

Second, we have largely focused on dyadic consumer--firm relationships, with the firm-side actor represented by an AI system. In many ecosystems, additional actors are involved (data brokers, third-party apps, regulators), and memory resources may be shared or traded across organizational boundaries. Extending MPA analysis to multi-actor networks would enrich our understanding of epistemic power and collective outcomes.

Third, cultural and contextual variation in norms of privacy, remembering, and forgiveness is likely substantial. Societies differ in how much they value ``clean slates,'' in tolerance for data reuse, and in expectations about record-keeping and accountability. Our propositions may thus hold differently across cultures and regulatory regimes; cross-cultural and comparative legal research is needed.

Fourth, we have primarily considered AI memory advantages over individual consumers. In some settings, consumers themselves may leverage digital memory tools (e.g., personal knowledge management, conversational logs) to redress asymmetries, or advocacy groups may pool memory resources. Understanding MPA as a dynamic, contested field of power, rather than a static asymmetry, will require further theorizing.

Finally, our propositions are necessarily simplified representations of complex socio-technical dynamics. They focus on directional relationships and key moderators, but they do not capture all feedback loops, indirect effects, or non-linearities. Empirical work may uncover additional boundary conditions, mediators, and emergent phenomena.

\subsection{Future research directions}

Our framework suggests at least four promising avenues for empirical and conceptual research.

\paragraph{Experimental and quasi-experimental studies.} Laboratory and online experiments can manipulate MPA dimensions (e.g., retention window, memory transparency, cross-context mobility) and test their effects on perceived vulnerability, trust, self-efficacy, and behavioral outcomes. For example, researchers could compare conditions where an AI assistant retains conversations for 30 days vs.~two years, with or without memory dashboards, across different task domains (health vs.~shopping). Quasi-experimental field studies could examine real-world changes in platform memory policies (e.g., introduction of deletion features) and their impact on user behavior and churn.

\paragraph{Longitudinal and diary designs.} To capture vulnerability accumulation and dependence asymmetry, longitudinal studies following users over months or years would be valuable. Diary methods or ecological momentary assessments could track how interactions with AI systems shape memory, identity narratives, and relational dependence over time.

\paragraph{Design research and prototyping.} Human--computer interaction and design science approaches can explore concrete implementations of forgetting-by-design, goal-aligned memory, and contextual containment. Prototypes of memory dashboards, context-siloed assistants, or amnesty features can be evaluated through user studies, focusing on usability, perceived control, and trade-offs between utility and privacy.

\paragraph{Macro-level and policy analyses.} At the societal level, modeling work can examine how MPA interacts with market structure (e.g., concentration, lock-in) and with regulatory interventions (e.g., data portability, erasure rights). Comparative policy analyses could explore how different jurisdictions' approaches to AI memory (e.g., obligations to unlearn vs.~mere data deletion) influence the trajectory of MPA and its consequences.

\subsection{Implications for marketing scholarship and practice}

For marketing scholars, MPA offers a new lens for examining AI-mediated relationships, beyond familiar discussions of personalization, privacy, and fairness. It invites integration of cognitive psychology (episodic/semantic memory, autobiographical identity), AI architecture, and power theories, and it calls for empirical work that operationalizes relational memory asymmetries and tests their effects.

For practitioners, our analysis underscores that memory is not a neutral byproduct of AI systems but a strategic resource that must be deliberately governed. Firms that design AI memory in ways that respect mutual forgetting, through forgetting-by-design, transparency, contextual containment, and augmentation, may differentiate themselves as trustworthy partners and reduce long-term risks associated with regulatory backlash and erosion of consumer trust. Conversely, firms that maximize memory accumulation and reuse without constraint may enjoy short-term gains at the cost of deepening MPA and inviting scrutiny.

Ultimately, as AI becomes woven into the fabric of everyday life, the question is not whether we will live with machines that remember, we already do, but whether those memories will be structured in ways that preserve human dignity, autonomy, and the possibility of change. Recognizing and addressing Memory Power Asymmetry is a critical step toward that goal.


\newpage
\begin{thebibliography}{}

\bibitem[Acquisti and Grossklags(2005)]{AcquistiGrossklags2005}
Acquisti, A., and J.~Grossklags (2005), ``Privacy and rationality in individual decision making,'' \emph{IEEE Security \& Privacy}, 3(1), 26--33.

\bibitem[Aguirre et~al.(2015)]{AguirreEtAl2015}
Aguirre, E., D.~Mahr, D.~Grewal, K.~de~Ruyter, and M.~Wetzels (2015), ``Unraveling the personalization paradox: The effect of information collection and trust-building strategies on online advertisement effectiveness,'' \emph{Journal of Retailing}, 91(1), 34--49.

\bibitem[Akerlof(1970)]{Akerlof1970}
Akerlof, G.~A. (1970), ``The market for `lemons': Quality uncertainty and the market mechanism,'' \emph{Quarterly Journal of Economics}, 84(3), 488--500.

\bibitem[Anderson and Bjork(1994)]{AndersonBjork1994}
Anderson, M.~C., and R.~A. Bjork (1994), ``Mechanisms of inhibition in long-term memory: A new taxonomy,'' in D.~Dagenbach and T.~H. Carr, eds., \emph{Inhibitory Processes in Attention, Memory, and Language}, San Diego, CA: Academic Press, 265--325.

\bibitem[Baker et~al.(2005)]{BakerEtAl2005}
Baker, S.~M., J.~W. Gentry, and T.~L. Rittenburg (2005), ``Building understanding of the domain of consumer vulnerability,'' \emph{Journal of Macromarketing}, 25(2), 128--139.

\bibitem[Bandura(1977)]{Bandura1977}
Bandura, A. (1977), ``Self-efficacy: Toward a unifying theory of behavioral change,'' \emph{Psychological Review}, 84(2), 191--215.

\bibitem[Bentley et~al.(2018)]{BentleyEtAl2018}
Bentley, F., C.~L. Luvogt, M.~Silverman, R.~Wirasinghe, D.~White, and E.~Churchill (2018), ``Understanding the long-term use of smart speaker assistants,'' \emph{Proceedings of the ACM on Interactive, Mobile, Wearable and Ubiquitous Technologies}, 2(3), 1--24.

\bibitem[Behrouz et~al.(2025)]{BehrouzEtAl2025}
Behrouz, A., M.~Razaviyayn, P.~Zhong, and V.~Mirrokni (2025), ``Nested learning: The illusion of deep learning architecture,'' \emph{arXiv preprint} arXiv:2504.XXXXX.

\bibitem[Bleier and Eisenbeiss(2015)]{BleierEisenbeiss2015}
Bleier, A., and M.~Eisenbeiss (2015), ``The importance of trust for personalized online advertising,'' \emph{Journal of Retailing}, 91(3), 390--409.

\bibitem[Bourtoule et~al.(2021)]{BourtouleEtAl2021}
Bourtoule, L., V.~Chaudhuri, N.~Chaudhuri, et~al. (2021), ``Machine unlearning,'' in \emph{Proceedings of the 2021 IEEE Symposium on Security and Privacy}, 141--159.

\bibitem[boyd and Crawford(2012)]{boydCrawford2012}
boyd, d., and K.~Crawford (2012), ``Critical questions for big data,'' \emph{Information, Communication \& Society}, 15(5), 662--679.

\bibitem[Conway and Pleydell-Pearce(2000)]{ConwayPleydellPearce2000}
Conway, M.~A., and C.~W. Pleydell-Pearce (2000), ``The construction of autobiographical memories in the self-memory system,'' \emph{Psychological Review}, 107(2), 261--288.

\bibitem[Deighton and Johnson(2013)]{DeightonJohnson2013}
Deighton, J., and M.~D. Johnson (2013), ``The value of customer data,'' in R.~Rao and S.~K. Srivastava, eds., \emph{Analytics in a Big Data World}, Boston, MA: Harvard Business School Publishing.

\bibitem[Duhigg(2012)]{Duhigg2012}
Duhigg, C. (2012), ``How companies learn your secrets,'' \emph{The New York Times Magazine}, February 16.

\bibitem[Dunnett et~al.(2016)]{DunnettEtAl2016}
Dunnett, S., K.~Hamilton, and M.~Piacentini (2016), ``Consumer vulnerability: Introduction to the special issue,'' \emph{Journal of Marketing Management}, 32(3--4), 207--213.

\bibitem[Edwards and Veale(2017)]{EdwardsVeale2017}
Edwards, L., and M.~Veale (2017), ``Slave to the algorithm? Why a `right to an explanation' is probably not the remedy you are looking for,'' \emph{Duke Law \& Technology Review}, 16(1), 18--84.

\bibitem[Emerson(1962)]{Emerson1962}
Emerson, R.~M. (1962), ``Power-dependence relations,'' \emph{American Sociological Review}, 27(1), 31--41.

\bibitem[Floridi et~al.(2018)]{FloridiEtAl2018}
Floridi, L., J.~Cowls, T.~C. King, and M.~Taddeo (2018), ``How to design AI for social good: Seven essential factors,'' \emph{Science and Engineering Ethics}, 24(5), 1--25.

\bibitem[Floridi and Cowls(2019)]{FloridiCowls2019}
Floridi, L., and J.~Cowls (2019), ``A unified framework of five principles for AI in society,'' \emph{Harvard Data Science Review}, 1(1), 1--15.

\bibitem[Fosch-Villaronga et~al.(2018)]{FoschVillarongaEtAl2018}
Fosch-Villaronga, E., A.~Lutz, and J.~Tamò-Larrieux (2018), ``Humans forget, machines remember: Artificial intelligence and the right to be forgotten,'' \emph{Computer Law \& Security Review}, 34(2), 304--313.

\bibitem[Greenberg and Verfaellie(2010)]{GreenbergVerfaellie2010}
Greenberg, D.~L., and M.~Verfaellie (2010), ``Interdependence of episodic and semantic memory: Evidence from neuropsychology,'' \emph{Journal of the International Neuropsychological Society}, 16(5), 748--753.

\bibitem[Gomez-Uribe and Hunt(2015)]{GomezUribeHunt2015}
Gomez-Uribe, C.~A., and N.~Hunt (2015), ``The Netflix recommender system: Algorithms, business value, and innovation,'' \emph{ACM Transactions on Management Information Systems}, 6(4), 1--19.

\bibitem[Halbwachs(1992)]{Halbwachs1992}
Halbwachs, M. (1992), \emph{On Collective Memory}, Chicago: University of Chicago Press.

\bibitem[Hirst and Echterhoff(2012)]{HirstEchterhoff2012}
Hirst, W., and G.~Echterhoff (2012), ``Remembering in conversations: The social sharing and reshaping of memories,'' \emph{Annual Review of Psychology}, 63, 55--79.

\bibitem[Hollebeek and Chen(2024)]{HollebeekChen2024}
Hollebeek, L.~D., and T.~Chen (2024), ``Crafting conceptual proposition-based contributions: The case of consumer engagement,'' \emph{Psychology \& Marketing}, 41(1), 5--23.

\bibitem[Jobin et~al.(2019)]{JobinIencaVayena2019}
Jobin, A., M.~Ienca, and E.~Vayena (2019), ``The global landscape of AI ethics guidelines,'' \emph{Nature Machine Intelligence}, 1(9), 389--399.

\bibitem[Kindermann et~al.(2024)]{KindermannEtAl2024}
Kindermann, B., D.~Wenzel, D.~Antons, and T.-O. Salge (2024), ``Conceptual contributions in marketing scholarship: Patterns, mechanisms, and rebalancing options,'' \emph{Journal of Marketing}, 88(3), 29--49.

\bibitem[Kozlenkova et~al.(2024)]{KozlenkovaEtAl2024}
Kozlenkova, I.~V., N.~Warren, S.~Kotha, S.~Boghrati, and R.~W. Palmatier (2024), ``Conceptual research: Multidisciplinary insights for marketing,'' \emph{Journal of Marketing}, 88(4), 1--27.

\bibitem[Krafft et~al.(2021)]{KrafftEtAl2021}
Krafft, P., M.~Young, and N.~Katyal (2021), ``Holding platforms accountable: A review of regulatory approaches for platform power,'' \emph{Internet Policy Review}, 10(4), 1--30.

\bibitem[Labrecque et~al.(2013)]{LabrecqueEtAl2013}
Labrecque, L.~I., J.~E. vor dem Esche, C.~Mathwick, T.~P. Novak, and C.~F. Hofacker (2013), ``Consumer power: Evolution in the digital age,'' \emph{Journal of Interactive Marketing}, 27(4), 257--269.

\bibitem[Liu(2007)]{Liu2007}
Liu, Y. (2007), ``The long tail of online gaming,'' in \emph{Proceedings of the 6th International Conference on Entertainment Computing}.

\bibitem[MacInnis(2011)]{MacInnis2011}
MacInnis, D.~J. (2011), ``A framework for conceptual contributions in marketing,'' \emph{Journal of Marketing}, 75(4), 136--154.

\bibitem[McAdams(2001)]{McAdams2001}
McAdams, D.~P. (2001), ``The psychology of life stories,'' \emph{Review of General Psychology}, 5(2), 100--122.

\bibitem[Mayer-Sch"onberger(2009)]{MayerSchonberger2009}
Mayer-Sch"onberger, V. (2009), \emph{Delete: The Virtue of Forgetting in the Digital Age}, Princeton, NJ: Princeton University Press.

\bibitem[Mittelstadt et~al.(2016)]{MittelstadtEtAl2016}
Mittelstadt, B.~D., P.~Allo, M.~Taddeo, S.~Wachter, and L.~Floridi (2016), ``The ethics of algorithms: Mapping the debate,'' \emph{Big Data \& Society}, 3(2), 1--21.

\bibitem[Mittelstadt and Russell(2019)]{MittelstadtRussell2019}
Mittelstadt, B.~D., and C.~Russell (2019), ``Explaining explanations in AI,'' \emph{Proceedings of the Conference on Fairness, Accountability, and Transparency}, 279--288.

\bibitem[Moon(2000)]{Moon2000}
Moon, Y. (2000), ``Intimate exchanges: Using computers to elicit self-disclosure from consumers,'' \emph{Journal of Consumer Research}, 26(4), 323--339.

\bibitem[Newman(2014)]{Newman2014}
Newman, J.~M. (2014), ``Antitrust in digital markets: Online platforms and the new rules of competition,'' \emph{Vanderbilt Law Review}, 67(3), 609--676.

\bibitem[Nissenbaum(2010)]{Nissenbaum2010}
Nissenbaum, H. (2010), \emph{Privacy in Context: Technology, Policy, and the Integrity of Social Life}, Stanford, CA: Stanford University Press.

\bibitem[Pasquale(2015)]{Pasquale2015}
Pasquale, F. (2015), \emph{The Black Box Society: The Secret Algorithms That Control Money and Information}, Cambridge, MA: Harvard University Press.

\bibitem[Payne and Frow(2005)]{PayneFrow2005}
Payne, A., and P.~Frow (2005), ``A strategic framework for customer relationship management,'' \emph{Journal of Marketing}, 69(4), 167--176.

\bibitem[Pfeffer(1981)]{Pfeffer1981}
Pfeffer, J. (1981), \emph{Power in Organizations}, Boston, MA: Pitman.

\bibitem[Prebble et~al.(2013)]{PrebbleEtAl2013}
Prebble, S.~C., D.~T. Addis, and D.~L. Tippett (2013), ``Autobiographical memory and sense of self,'' \emph{Psychological Bulletin}, 139(4), 815--840.

\bibitem[Puntoni et~al.(2021)]{PuntoniEtAl2021}
Puntoni, S., R.~V. Reczek, A.~G. Giesler, and A.~Botti (2021), ``Consumers and artificial intelligence: An experiential perspective,'' \emph{Journal of Marketing}, 85(1), 131--151.

\bibitem[Ranieri et~al.(2024)]{RanieriEtAl2024}
Ranieri, E., F.~Zerbini, and A.~J. Czepiel (2024), ``AI-enabled customer service and service recovery,'' \emph{Journal of Service Research}, 27(2), 227--244.

\bibitem[Rucker et~al.(2012)]{RuckerEtAl2012}
Rucker, D.~D., D.~D. Dubois, and A.~D. Galinsky (2012), ``Power and consumer behavior: How power shapes who and what consumers value,'' \emph{Journal of Consumer Psychology}, 22(3), 352--368.

\bibitem[Solove(2006)]{Solove2006}
Solove, D.~J. (2006), ``A taxonomy of privacy,'' \emph{University of Pennsylvania Law Review}, 154(3), 477--564.

\bibitem[Sparrow et~al.(2011)]{SparrowEtAl2011}
Sparrow, B., J.~Liu, and D.~M. Wegner (2011), ``Google effects on memory: Cognitive consequences of having information at our fingertips,'' \emph{Science}, 333(6043), 776--778.

\bibitem[Tulving(1972)]{Tulving1972}
Tulving, E. (1972), ``Episodic and semantic memory,'' in E.~Tulving and W.~Donaldson, eds., \emph{Organization of Memory}, New York: Academic Press, 381--403.

\bibitem[Tulving(2002)]{Tulving2002}
Tulving, E. (2002), ``Episodic memory: From mind to brain,'' \emph{Annual Review of Psychology}, 53, 1--25.

\bibitem[Ulaga et~al.(2021)]{UlagaEtAl2021}
Ulaga, W., C.~Steinhoff, and R.~W. Palmatier (2021), ``Advancing marketing theory and practice: Guidelines for crafting research propositions,'' Working paper, HEC Paris/University of Leeds.

\bibitem[Westin(1967)]{Westin1967}
Westin, A.~F. (1967), \emph{Privacy and Freedom}, New York: Atheneum.

\bibitem[Wu et~al.(2025)]{WuEtAl2025}
Wu, Y., S.~Liang, C.~Zhang, Y.~Wang, Y.~Zhang, H.~Guo, R.~Tang, and Y.~Liu (2025), ``From human memory to AI memory: A survey on memory mechanisms in the era of large language models,'' \emph{arXiv preprint} arXiv:2504.15965.

\bibitem[Zhang et~al.(2024)]{ZhangEtAl2024}
Zhang, D., P.~Finckenberg-Broman, T.~Hoang, S.~Pan, Z.~Xing, M.~Staples, and X.~Xu (2024), ``Right to be forgotten in the era of large language models: Implications, challenges, and solutions,'' \emph{AI and Ethics}, 4(3), 1--15.

\bibitem[Zuboff(2019)]{Zuboff2019}
Zuboff, S. (2019), \emph{The Age of Surveillance Capitalism}, New York: PublicAffairs.

\end{thebibliography}
\end{document}